\documentclass[twocolumn,groupedaddress,assymb,amsmath]{revtex4}
\usepackage{graphicx}
\usepackage{amssymb}
\usepackage{amsmath}

\usepackage{extarrows}
\usepackage{color}
\usepackage[colorlinks]{hyperref}
\usepackage{lipsum} % 调用跨栏长公式宏包

\begin{document}
\title{Detecting the effects of quantum gravity with exceptional points in optomechanical sensors}
  \author{Dianzhen Cui$^{1}$, T. Li$^{1}$, Jianning Li$^{1}$, Xuexi Yi$^{1,2}$\footnote{yixx@nenu.edu.cn}}
\affiliation{$^1$Center for Quantum Sciences and School of Physics, Northeast Normal University, Changchun 130024, China\\
$^2$Center for Advanced Optoelectronic Functional Materials Research, and Key Laboratory for UV Light-Emitting
{Materials and Technology of Ministry of Education, Northeast Normal University, Changchun 130024, China}}

  \date{\today}
 \begin{abstract}
In this manuscript, working with a binary mechanical system, we examine the effect of quantum gravity on the exceptional points of the system. On the one side, we find  that the exceedingly weak effect of quantum gravity can be sensed via pushing the system towards a second-order exceptional point, where the spectra of the non-Hermitian system exhibits non-analytic and even discontinuous behavior. On the other side,  the gravity perturbation will affect the sensitivity of the system to deposition mass. In order to further enhance the sensitivity of the system to quantum gravity, we  extend the system to the other one which has a higher-order (third-order) exceptional point. Our work provides a feasible way to use exceptional points as a new tool to explore the effect  of quantum gravity.
 \end{abstract}
 %\pacs{}
\maketitle

\section{Introduction}
Cavity optomechanics \cite{Aspelmeyer2014}, exploring the interaction between light and mechanical systems, has made a profound impact in recent years due to its wide variety of applications including optomechanical sensors. Optomechanical  sensors have achieved ultrasensitive performance in  gravitational wave detection \cite{Caves1980,Abramovici1992}, high-precision measurements \cite{Matsumoto2019}, detection for mass \cite{Liu2019}, acceleration \cite{Krause2012}, displacement \cite{Rossi2018}, and force \cite{C. M. Caves1980, Schreppler2014,Basiri-Esfahani2019}. For practical applications, the optomechanical system is unavoidably coupled with its surroundings, leading to a non-Hermitian optomechanical system. Several earlier studies have shown that non-Hermitian  spectral degeneracies, also known as exceptional points (EPs) \cite{Peng2014,Wiersig2014,Peng2016,Chen2017,Lv2018}, governs the dynamics of parity-time $(\mathcal{PT})$ symmetric system subject to environment. In contrast to level degeneracy points in Hermitian systems, the EP is associated with level coalescence, in which the eigenenergies and the corresponding eigenvectors simultaneously coalesce \cite{Heiss2004,Berry2004}. Besides, the intriguing phenomena of EPs in unidirectional invisibility \cite{Lin2011},  topological chirality \cite{Xu2016} and low-threshold lasers \cite{Jing2014,Feng2014} have been predicted.

In recent years, sensitivity enhancement of the sensor operating at  EPs has been explored both theoretically   \cite{Wiersig2014,Djorwe2019,Ren2017} and   experimentally \cite{Lai2019,Chen2017,Lai2020,Hokmabadi2019} in  a number of systems including  nanoparticle detector \cite{Wiersig2014,Chen2017}, mass sensor \cite{Djorwe2019}, and gyroscope \cite{Lai2019, Ren2017,Lai2020,Khial2018}. These studies have shown that if a second-order exceptional point (EP2) where the coalescence of two levels occurs is subjected to a perturbation of strength $\epsilon$, the frequency splitting (the energy spacing of the two levels) is typically proportional to the square root of the perturbation strength $\epsilon$. This is the so-called complex square-root topology. Moreover, the splitting is significantly enhanced for sufficiently small perturbation. This suggests  that the use of EP can enhance the sensitivity of a quantum sensor.

In standard quantum mechanics, on the basis of Heisenberg uncertainty principle $\Delta x \Delta p \geq \frac{\hbar}{2}$ \cite{Heisenberg1927}, the position $x$ and the momentum $p$ of an particle cannot be simultaneously measured to arbitrary precision, however, the uncertainty of $x$ can reach zero in case $\Delta p$ approaches  infinity.  This is not the case when the quantum gravity is taken into account. It has been  suggested  that the uncertainty relation should be modified when gravitational effects have been taken into consideration \cite{Garay1995}. Such generalized uncertainty principle (GUP) is found in various approaches to quantum gravity, such as the finite bandwidth approach to quantum gravity \cite{Kempf2009,Martin2012}, string theory \cite{Veneziano1986,Amati1987,Gross1988,Amati1989,Konishi1990}, the theory of double special relativity \cite{Amelino-Camelia2002,Magueijo2003}, relative locality \cite{Amelino-Camelia2011} and black holes \cite{Scardigli1999}.

The generalized Heisenberg uncertainty principle that counts the gravity effects is  $\Delta x \Delta p\geq \frac{\hbar}{2}(1+\mu \Delta p^2)$ \cite{Amati1989}. Here $\mu=\frac{\beta_0}{(M_p c)^2}=\frac{L_p^2}{2 \hbar^2}$, $\beta_0$ is a dimensionless parameter, $M_p$ is Planck mass, $M_p c^2$ is Planck energy and $L_p$ is Plank length. This inequality means that $\Delta x\geq L_p\sqrt{\beta_0}$. So if $\beta_0=1$ \cite{Das2008},  the minimal length is equal to the Planck length $(L_p)$ beyond which the concepts of time and space will lose their meaning. The generalized uncertainty principle (GUP) has been extensively explored  in various fields, including high energy physics, cosmology and black holes \cite{Zhong-Wen2017}. Due to experiments that can test minimal length scale directly require energies much higher than that currently available, most of the work has been devoted to find {\it indirect evidences} of quantum gravity in high energy particle collisions and astronomical observations \cite{Amelino-Camelia1998,Jacob2007}.

In this manuscript, we theoretically propose the other sensing scheme to explore the effects of quantum gravity via GUP. We will consider a binary and a ternary mechanical system separately  within an optomechanical configuration.  Controlling  the  gain and loss of the mechanical oscillators  and driving  the two cavities with blue and red detuned lasers as well as  manipulating  the strength of the electromagnetic field $(\alpha^{in})$, we can set the system into a self-sustained regime for the mechanical oscillations. In the absence of gravity perturbation, the system is controlled to be in the EP2 state. When the gravity perturbation comes, the supermodes of the system are shifted away from the EP2. The frequency splitting induced by gravitational effects can be read out in the mechanical spectrum. This result has been further extended to a third-order exceptional point (EP3) by taking a more complicated ternary mechanical system into account. Compared with the scheme utilizing EP2, optomechanical sensor based on EP3 performs better. The physics of this EP-based sensor is that the eigenvalues  of non-Hermitian Hamiltonian  may exhibit non-analytic and even discontinuous behavior, which in principle enables an unlimited spectral sensitivity.

The rest of this paper is organized as follows. In Sec.~\ref{sec1}, we introduce  the physical model and derive a set of equation for the  dynamics of our system. In Sec.~\ref{sec2}, we study the sensitivity of the system to gravity perturbation  at EP2. In Sec.~\ref{sec3}, we extend the study on the sensitivity of the system to the gravity perturbation to  EP3. In Sec.~\ref{sec4}, we discuss experimental feasibility of the proposed sensing scheme and analyze the  limitation  of the proposed quantum gravity sensor. Finally, the conclusions are drawn in Sec.~\ref{sec5}. In Appendix \ref{appendix}, we present details of derivation for the effective Hamiltonian.

\section{general framework}\label{sec1}
We start by briefly recalling  the description  of harmonic oscillator with mass $m$ when  the  effect of  gravity is taken into consideration. Afterwards we would apply this result to our model, in which the mechanical resonator is modeled as a harmonic oscillator.
  \subsection{Deformed harmonic oscillations under quantum gravity}

From the aspect of communtation relation, the gravity would modify the relation leading to  the generalized uncertainty principle (GUP) given in Ref.  \cite{Amati1989},
\begin{eqnarray}
\begin{aligned}
&[x,p]=i \hbar (1+\mu p ^2).
\label{commutation relation}
\end{aligned}
\end{eqnarray}
Define \cite{Das2008}
\begin{eqnarray}
\begin{aligned}
&p=(1+\frac{1}{3} \mu \widetilde{p}^2)\widetilde{p},
\label{map}
\end{aligned}
\end{eqnarray}
where $x$, $\widetilde{p}$ satisfying the (non deformed) canonical commutation relations $[x,\widetilde{p}]=i\hbar$. It is easy to see that Eq. (\ref{map}) is written up to the first  order in $\mu$ (the terms of order $\mu^2$ and higher are neglected). For a harmonic oscillator with  frequency $\omega$ and mass $m$, we assume that the Hamiltonian takes $H=\frac{1}{2}m \omega^2 x^2+\frac{p^2}{2m}$ \cite{Pikovski2012}. In terms of  $\widetilde{p}$,  the Hamiltonian of harmonic oscillator can  be rewritten as
\begin{eqnarray}
\begin{aligned}
H=\frac{1}{2}m \omega^2 x^2+\frac{\widetilde{p}^2}{2m}+\frac{\mu \widetilde{p}^4}{3m}.
\label{Hamiltonian of harmonic oscillator}
\end{aligned}
\end{eqnarray}
Introducing canonical creation and annihilation operators,
\begin{eqnarray}
\begin{aligned}
b^\dagger=\sqrt{\frac{m\omega}{2\hbar}}(x-\frac{i\widetilde{p}}{m\omega}), b=\sqrt{\frac{m\omega}{2\hbar}}(x+\frac{i\widetilde{p}}{m\omega})
\label{operators}
\end{aligned}
\end{eqnarray}
and rewriting the Hamiltonian (\ref{Hamiltonian of harmonic oscillator}) in terms of $b$,  we can get
\begin{eqnarray}
\begin{aligned}
H=&\hbar\omega(b^\dagger b+\frac{1}{2})+\frac{1}{12}\hbar^2 \omega^2 m \mu(b-b^\dagger)^4.
\label{Hamiltonian_Rewriting}
\end{aligned}
\end{eqnarray}
The first term represents the free Hamiltonian of the harmonic oscillator. The second term comes from the gravitational effects.

  %figure 1
\begin{figure}[h]
\vspace{-2.48cm}
	\centering

	\includegraphics[width=0.45\textwidth]{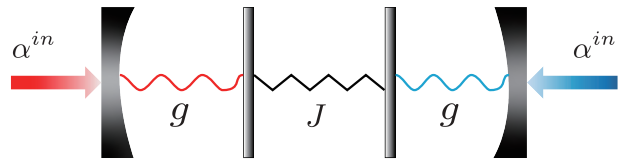}

	\caption {Sketch of the studied system. An open system consisted of two coupled mechanical resonators. Each resonator is coupled to an optical cavity. The two cavities are driven by red- and blue-detuned pump laser, respectively.}
	\label{setup1}
\vspace{-0.8cm}
\end{figure}

  \subsection{Modeling and dynamical equations}
We consider two coupled  identical  resonators with optomechanically induced gain and loss. Each of the resonators is characterized by   frequency $\omega_j$, damping rate $\gamma_m$ and  coupling  strength $J$.  The schematic diagram is presented in Fig. \ref{setup1}. In this configuration, the theory states  that only the mechanical commutation relation are modified, while the optical commutation relation remains unchanged, i.e.,  $[a,a^\dagger]=1$ \cite{Girdhar2020}.  The total Hamiltonian of the whole system  can be written as $(\hbar=1)$
\begin{eqnarray}
\begin{aligned}
H=H_{f}+H_{i}+H_{d}+H_{g},
\label{Hamiltonian of set up1}
\end{aligned}
\end{eqnarray}
where,
\begin{eqnarray}
\begin{aligned}
H_{f}&=\sum_{j=1,2}\omega_{a,j} a_j^\dagger a_j+\omega_j b_j^\dagger b_j,\\H_{i}&=\sum_{j=1,2}\big\{-g a_j^\dagger a_j(b_j^\dagger+b_j)\big\}-J(b_1b_2^\dagger+b_1^\dagger b_2),\\H_{d}&=\sum_{j=1,2}iE(a_j^\dagger e^{-i\omega_{p,j}t}-a_j e^{i\omega_{p,j}t}),\\H_{g}&=\sum_{j=1,2}\frac{1}{12} \omega_j^2 m_j \mu_j(b_j-b_j^\dagger)^4.
\label{Hamiltonian of set up1_expend}
\end{aligned}
\end{eqnarray}

In this expression, $H_{f}$ represents the sum of free Hamiltonian of the optomechanical system, $a_j^\dagger$ $(b_j^\dagger)$ and $a_j$ $(b_j)$ are the  creation and annihilation operators of the $j$th cavity (mechanical resonator) $(j = 1, 2)$. The frequencies of the cavities and mechanical resonators are $\omega_{a,j}$ and $\omega_j$, respectively. $H_{i}$ describes the interaction Hamiltonian of the configuration. The first term represents the coupling of the cavities to the corresponding mechanical resonators with optomechanical coupling strength $g$. The second term describes the coupling between the two mechanical resonators with coupling strength $J$. $H_{d}$ indicates that the two cavities are driven by external fields with amplitude $E$ and frequency $\omega_{p,j}$. $H_{g}$ describes the  gravitational effects in  mechanical resonators. The effective mass of the $j$th mechanical mode is $m_j$. In the frame rotating at the input laser frequency $\omega_p$, the Hamiltonian of the system reads,
\begin{eqnarray}
\begin{aligned}
H=&\sum_{j=1,2} \Big\{-\Delta_j a_j^\dagger a_j+\omega_j b_j^\dagger b_j-g a_j^\dagger a_j(b_j^\dagger+b_j)\\&+\frac{1}{12} \omega_j^2 m_j \mu_j(b_j-b_j^\dagger)^4+iE(a_j^\dagger-a_j)\Big\}
\\&-J(b_1b_2^\dagger+b_1^\dagger b_2),
\label{Hamiltonian of set up_rotating}
\end{aligned}
\end{eqnarray}
where, $\Delta_j=\omega_{p,j}-\omega_{a,j}$ represents the detuning of the driving field with respect to the cavity. As we are interested in the classical limit, where photon and phonon numbers are assumed large in the model. Thus, we replace the quantum operators with their mean values, i.e., $\alpha_j=\left\langle a \right\rangle$ and $\beta_j=\left\langle b \right\rangle$. By introducing dissipation terms, the evolution of the system operators  is obtained as follows \cite{Aspelmeyer2014},
\begin{eqnarray}
\begin{aligned}
\frac{d\alpha_j}{dt}=&[i(\Delta_j+g(\beta_j^*+\beta_j))-\frac{\kappa}{2}]\alpha_j+\sqrt{\kappa}\alpha_j^{in},
\\ \frac{d\beta_j}{dt}=&-(i\omega_j+\frac{\gamma_m}{2})\beta_j+iJ\beta_{3-j}+ig\alpha_j^* \alpha_j\\&+\frac{1}{3}i m_j\omega_j^2\mu_j(\beta_j-\beta_j^*)^3,
\label{QLEs1}
\end{aligned}
\end{eqnarray}
where $\kappa$ and $\gamma_m$ are the intrinsic damping rates of the cavities and mechanical resonators, respectively. $E=\sqrt{\kappa}\alpha_j^{in}$ is the amplitude of the driving field, where $\alpha_j^{in}=\sqrt{\frac{p_{in}}{\hbar \omega_{p,j}}}$ characterizes the input field driving the cavity. For the sake of simplicity, we assume the two cavities and mechanical resonators identical, this means $\omega_j=\omega_m$, $m_j=m$, and $\mu_j=\mu$. We apply the input lasers with the same power $(p_{in})$ to drive the two mechanical resonators, i.e.,  $\alpha_j^{in}=\alpha^{in}$. Throughout the work, the parameters satisfy the following condition, $\gamma_m,g\ll\kappa\ll\omega_m$, similar to those chosen in Ref. \cite{Cohen2015,Hong2017}. Under this  hierarchy, the amplitude and phase of the mechanical resonators slowly evolving on the time scale of the cavity dynamics.

We will pay our attention to the steady state of the mechanical resonators. In this regime, $\beta_j(t)=\bar{\beta}_j +B_j e^{-i\theta}e^{-i\omega_{l} t}$ \cite{Marquardt2006,Rodrigues2010}, where $\bar{\beta}_j$ is the center of the mechanical oscillations and   amplitude $B_j$ can be regarded as a slowly evolving function of time. In the limit-cycle states, both mechanical resonators start oscillating with a locked frequency $(\omega_{l})$. On this point, it can be seen from its Fourier spectrum, where the peak of the spectrum is much larger than the corresponding amplitude of other frequency components \cite{Djorwe2018}. Throughout this paper, we set $\theta=0$. In parallel, we removed all terms in mechanical dynamics except for the constant one and the term oscillating at $\omega_l$. Using this analytic approximation, we solve the equation for $\alpha_j$ assuming a fixed mechanical amplitude and then substitute the result into the equation for $\beta_j$, resulting in the following set of equations of motion describing this effective mechanical system (see Appendix \ref{appendix}):
\begin{small}
\begin{eqnarray}
\begin{aligned}
&\frac{d\beta_1}{dt}=-(i\omega_{eff}^1+\frac{\gamma_{eff}^1}{2}+i \Theta_1)\beta_1+iJ\beta_2+i \Theta_1 \beta_1^*,
\\&\frac{d\beta_2}{dt}=-(i\omega_{eff}^2+\frac{\gamma_{eff}^2}{2}+i \Theta_2)\beta_2+iJ\beta_1+i \Theta_2 \beta_2^*,
%\\&\frac{d\beta_1^*}{dt}=-(-i\omega_{eff}^1+\frac{\gamma_{eff}^1}{2})\beta_1^*-iJ\beta_2^*-i\Theta_1(-\beta_1^*+\beta_1),
%\\&\frac{d\beta_2^*}{dt}=-(-i\omega_{eff}^2+\frac{\gamma_{eff}^2}{2})\beta_2^*-iJ\beta_1^*-i\Theta_2(-\beta_2^*+\beta_2),
\label{QLEs2}
\end{aligned}
\end{eqnarray}
\end{small}
where, $\Theta_j= \mu m \omega_j^2 B_j^2$ $(j=1,2)$. $\omega_{eff}^j=\omega_j+\Omega_j$ and $\gamma_{eff}^j=\gamma_m+\Gamma_j$ $(j=1,2)$ represent the effective frequency and damping of the $j$th mechanical oscillator $(j = 1, 2)$, respectively. The modal field evolution in this configuration obeys $i\frac{d\phi}{dt}=H_{eff}\phi$, where $\phi=(\beta_1,\beta_2,\beta_1^*,\beta_2^*)^T$ is the state vector and $t$ represents time. $H_{eff}$ is the associated $4\times4$ non-Hermitian Hamiltonian (see more details in Appendix \ref{appendix}):
\begin{small}
\begin{widetext}
\begin{equation}
H_{eff}= \left(
 \begin{array}{cccc}
 \omega_{eff}^1-i\frac{\gamma_{eff}^1}{2}+\Theta_1&-J&-\Theta_1&0\\
 -J&\omega_{eff}^2-i\frac{\gamma_{eff}^2}{2}+\Theta_2&0&-\Theta_2\\
 \Theta_1&0&-\omega_{eff}^1-i\frac{\gamma_{eff}^1}{2}-\Theta_1&J\\
 0&\Theta_2&J&-\omega_{eff}^2-i\frac{\gamma_{eff}^2}{2}-\Theta_2
 \end{array}
\right).
\end{equation}
\end{widetext}
\end{small}
Here, $\Omega_j$ $(\Gamma_j)$ represents the optical spring effect (the optomechanical damping rate). These quantities are given as (see more details in Appendix \ref{appendix})
\begin{eqnarray}
\begin{aligned}
\Omega_j&=-\frac{2\kappa(g\alpha^{in})^2}{\omega_{l} \epsilon_j} Re\left(\sum_{n}\frac{J_{n+1}(-\epsilon_j)J_{n}(-\epsilon_j)}{K_{n+1}^{j*}K_n^j}\right),
\label{optical spring effect}
\end{aligned}
\end{eqnarray}
and
\begin{eqnarray}
\begin{aligned}
\Gamma_j&=\frac{2(g\kappa\alpha^{in})^2}{\epsilon_j} \sum_{n}\frac{J_{n+1}(-\epsilon_j)J_{n}(-\epsilon_j)}{\left|K_{n+1}^{j*}K_n^j\right|^2}.
\label{optomechanical damping rate}
\end{aligned}
\end{eqnarray}

\begin{figure}[h]
	\centering
	\includegraphics[width=0.50\textwidth]{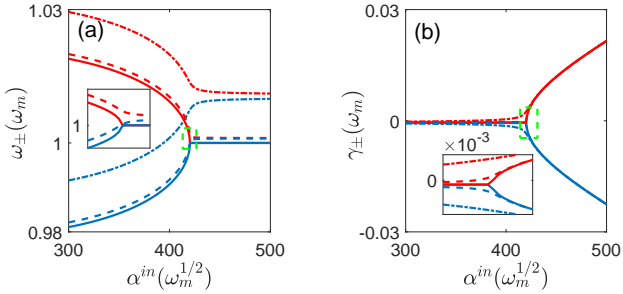}
	\caption {(a) The real and (b) the imaginary parts of the eigenvalues $vs$ the driving strength $\alpha^{in}$ before and after the perturbation introduced by gravitational effects. The system exhibits an EP2 at $4J=\gamma_{eff}^2-\gamma_{eff}^1$. The green dashed box is zoomed out in the inset to show the influence of gravitational effects on EP2. Each panel contains various curves for comparision: the solid lines denote the eigenvalues without gravitational effect, the dashed lines indicate the eigenvalues at $\mu m=(0.02\times2.2\times10^{-7})\omega_m^{-1}$, and the dash-dotted lines represent the eigenvalues at $\mu m=(0.02\times2.2\times10^{-6})\omega_m^{-1}$. The other parameters are chosen as  $J=2.2\times10^{-2}\omega_m$, $g=2.5\times10^{-4}\omega_m$, $\gamma_m=10^{-3}\omega_m$, $\kappa=10^{-1}\omega_m$, $\omega1=\omega2=\omega_m$, $\Delta1=-\omega_m$, and $\Delta2=\omega_m$. The colors (red, blue) represent a pair of eigenvalues of our model.}
	\label{twostate_benzhengzhi}
\end{figure}

Firstly, we focus on the case without gravitational effect, the eigenvalues of the above effective Hamiltonian are given by
\begin{eqnarray}
\begin{aligned}
\lambda_{\pm}=&\omega_l-\frac{i}{4} (\gamma_{eff}^1+\gamma_{eff}^2)\pm \frac{1}{4}\delta,
\label{supermodes}
\end{aligned}
\end{eqnarray}
where,
\begin{eqnarray}
\begin{aligned}
\delta=\sqrt{16J^2+[2(\omega_{eff}^1-\omega_{eff}^2)+i(\gamma_{eff}^2-\gamma_{eff}^1)]^2}.
\label{delta}
\end{aligned}
\end{eqnarray}
Here, $\omega_l=\frac{\omega_{eff}^1+\omega_{eff}^2}{2}$. Replace the conventional vibrational modes, we now have new mechanical modes, which can be called as the mechanical supermodes. The effective frequencies and spectral linewidths of the system are defined as the real $(\omega)$ and imaginary $(\gamma)$ parts of eigenvalues, respectively. The solid lines in Fig. \ref{twostate_benzhengzhi} show the real and imaginary parts of the eigenvalues $vs$ the driving strength $\alpha^{in}$ before the perturbation introduced by gravity effects. At the specific point, both these pairs of effective frequencies and effective dampings of the system coalesce.
%figure 2
It is evident that for a critical driving strength $\alpha^{in}$ the pairs of eigenvalues merge at $4J=\gamma_{eff}^2-\gamma_{eff}^1.$

\section{Sensitivity At The Second-Order Exceptional Point}\label{sec2}
\subsection{Sensitivity of a system at the second-order exceptional point to the gravity effect}
For the case with gravitational effects, we numerically solve the eigenvalues of this mechanical effective Hamiltonian and show the results in Fig. \ref{twostate_benzhengzhi}. The effective Hamiltonian has 4 eigenvalues forming two pairs, one pair is due to the apperarnace of $\beta_j^*$ in the dynamics.

As shown in Fig. \ref{twostate_benzhengzhi}, we see that the splitting of effective frequency (real part of the eigenvalue) and linewidth (imaginary part of the eigenvalue)  increases as the mass of the mechanical resonators increases. This is attributed to the fact that
gravity effect is enhanced by larger system mass. A typical detection strategy is to observe the associated mode response, usually the frequency splitting or the frequency shift, before and after the perturbation induced by gravitational effects taking place. In this paper,  in order to quantify the frequency splitting  caused by  the gravity effect, we define the sensitivity  as follows,
\begin{eqnarray}
\begin{aligned}
\Delta\omega=\left|Re\lambda_+-Re\lambda_-\right|.
\label{sensitivity}
\end{aligned}
\end{eqnarray}
The perturbation of gravitational effects can shift the EP2, and thereby the degeneracy of the effective frequencies are released and cause the supermodes to split.
%figure 3
\begin{figure}[h]
	\centering
	\includegraphics[width=0.5\textwidth]{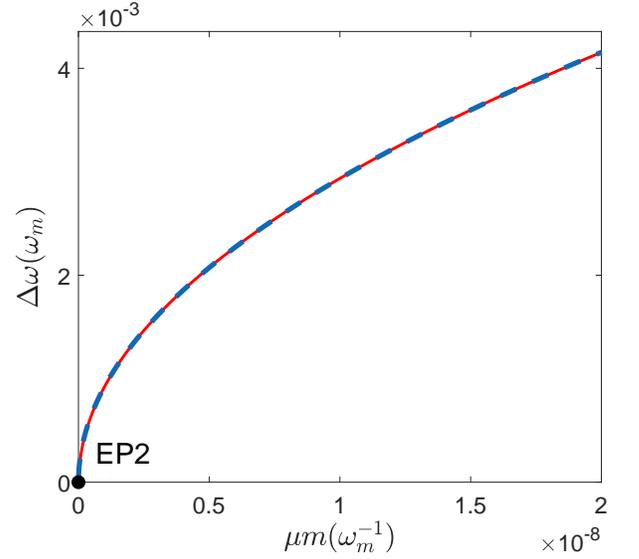}
	\caption {Sensitivity (red solid line) at the second-order EP $vs$ $\mu m$. The blue dashed line denotes the fitted curve with $\xi = 30.12\omega_m^{3/2}$. The parameters are the same as that in Fig. \ref{twostate_benzhengzhi}.}
	\label{yinli_lingmindu_EP}
\end{figure}
The frequency splitting caused by gravitational effects can be fitted using
\begin{eqnarray}
\begin{aligned}
\Delta\omega\approx\xi (\mu m)^{1/2}.
\label{fit_EP2}
\end{aligned}
\end{eqnarray}
Here, $\xi$ is the fitting coefficient. Fig. \ref{yinli_lingmindu_EP} shows $\Delta\omega$ as a function of the $\mu m$ near the EP2. The blue dashed lines represent the fitting result according to Eq. (\ref{fit_EP2}) with $\xi = 30.12\omega_m^{3/2}$, which is consistent with the mechanical frequency splitting in our model. Therefore, it can be inferred that the mechanical frequency splitting in response to the $\mu m$ obeys the square root behavior. Due to the intrinsic properties of EP2, we can claim  that the sensitivity is significantly enhanced by exploiting EP2 for sufficiently small perturbation strength, proving the efficiency of the EP2 sensor in detecting gravity effect.

\subsection{Sensitivity of the system to deposition mass at the second-order exceptional point with gravity effect}
In order to gain insight into the influence of gravity effects on mass sensing, we assume that a mass $\delta m$ has been deposited on the mechanical oscillator driven by the blue-detuned electromagnetic field, which would induce the frequency shift given in Eq. (\ref{supermodes}), i.e., replacing $\omega_{eff}^2$ with $\omega_{eff}^2+\delta\omega$. For an ordinary mass sensor, the relation  between the deposited mass $\delta m$ and the frequency shift $\delta\omega$ is given by \cite{Li2007}
\begin{eqnarray}
\begin{aligned}
\delta m=\frac{2m}{\omega_m} \delta \omega=\zeta^{-1} \delta \omega,
\label{ordinary mass sensor}
\end{aligned}
\end{eqnarray}
where $\zeta$ represents the mass responsivity of the mechanical resonator. We can define the gap as
\begin{eqnarray}
\begin{aligned}
\chi_{\pm}=\lambda_{\pm}(m)-\lambda_{\pm}.
\label{gap}
\end{aligned}
\end{eqnarray}
    %figure 4
\begin{figure}[h]
	\centering
    \includegraphics[width=0.50\textwidth]{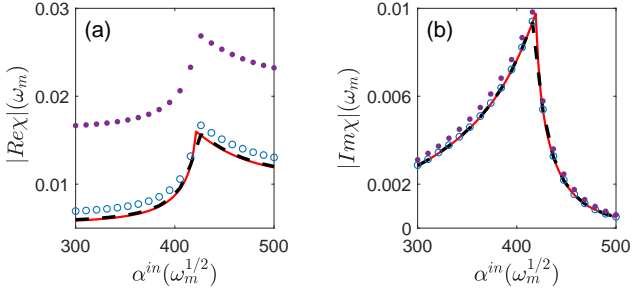}
	\caption {(a) The real and (b) the imaginary parts of $\chi$ $vs$ $\alpha^{in}$ before and after the perturbation induced by gravitational effects with frequency shift $\delta\omega=10^{-2}\omega_m$. Each panel contains various lines for comparision. The red solid lines denote the gap without gravitational effect, the black dashed lines indicate the gap at $\mu m=(0.02\times2.2\times10^{-8}) \omega_m^{-1}$, the blue circles represent the eigenvalues at $\mu m=(0.02\times2.2\times10^{-7})$ $\omega_m^{-1}$, and the purple filled circles mean the eigenvalues at $\mu m=(0.02\times2.2\times10^{-6}) \omega_m^{-1}$. Other system parameters are the same as that in Fig. \ref{twostate_benzhengzhi}. Note these results are for $\chi_+$.}
	\label{zhiliang_benzhengzhi}
\end{figure}
Figure \ref{zhiliang_benzhengzhi} shows that for mechanical frequency shift $\delta\omega=10^{-2}\omega_m$, the larger the mass of the mechanical resonators, the larger the gap between effective frequencies before and after gravitational effects being considered. However, the gap between the effective dampings (the imaginary part of the eigenvaules) does not change significantly. Therefore, small  mass of the mechanical resonators can decrease  the disturbance caused by gravitational effects.

\section{Sensitivity Of a system  At The Third-Order Exceptional Point to the   Gravity effect}\label{sec3}

\begin{figure}[h]
\vspace{-2.5cm}
	\centering
    \includegraphics[width=7cm,height=8cm]{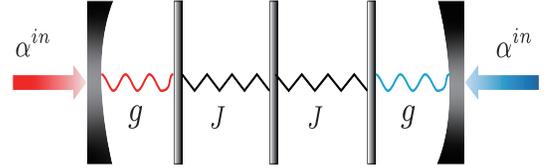}
\vspace{-2.5cm}
	\caption {A schematic diagram of model. An system consisted of coupled ternary resonators. }
	\label{setup2}
\end{figure}

Inspired by these results, we now extend this scheme  to the higher-order exceptional points (EPs). A possible configuration that supports a third-order exceptional point (EP3) would be a system consisting of two cavities and three coupled mechanical oscillators where the two cavities are symmetrically driven by red- and blue-detuned lasers, and the corresponding mechanical resonators are coupled together (see Fig. \ref{setup2}).
   %figure 5
Proceeding in a similar way, one can write the following the Hamiltonian of the system,
\begin{eqnarray}
\begin{aligned}
H=H_f+H_i+H_d+H_g,
\label{Hamiltonian of set up 2}
\end{aligned}
\end{eqnarray}
with
\begin{eqnarray}
\begin{aligned}
H_f=&\sum_{j=1,2}-\Delta_j a_j^\dagger a_j+\sum_{j=1,2,3}\omega_j b_j^\dagger b_j,
\\H_i=&-g a_1^\dagger a_1(b_1^\dagger+b_1)-g a_2^\dagger a_2(b_3^\dagger+b_3)\\&-J(b_1b_2^\dagger+b_1^\dagger b_2)-J(b_2b_3^\dagger+b_2^\dagger b_3),
\\H_d=&\sum_{j=1,2}iE(a_j^\dagger-a_j),
\\H_g=&\sum_{j=1,2,3}\frac{1}{12} \omega_j^2 m_j \mu_j(b_j-b_j^\dagger)^4.
\label{Hamiltonian of set up 2_expend}
\end{aligned}
\end{eqnarray}
From Eq. (\ref{Hamiltonian of set up 2_expend}), one can write the following nonlinear equations of motion,
\begin{small}
\begin{eqnarray}
\begin{aligned}
&\frac{d\alpha_j}{dt}=[i(\Delta_j+g(\beta_j^*+\beta_j))-\frac{\kappa}{2}]\alpha_j+\sqrt{\kappa}\alpha_j^{in},
\\&\frac{d\beta_1}{dt}=-(i\omega_1+\frac{\gamma_m}{2})\beta_1+iJ\beta_{2}+ig\alpha_1^* \alpha_1+\Xi_1(\beta_1-\beta_1^*)^3,
\\&\frac{d\beta_2}{dt}=-(i\omega_2+\frac{\gamma_m}{2})\beta_2+iJ\beta_{1}+iJ\beta_{3}+\Xi_2(\beta_2-\beta_2^*)^3,
\\&\frac{d\beta_3}{dt}=-(i\omega_3+\frac{\gamma_m}{2})\beta_3+iJ\beta_{2}+ig\alpha_2^* \alpha_2+\Xi_3(\beta_3-\beta_3^*)^3.
\label{QLEs-EP3}
\end{aligned}
\end{eqnarray}
\end{small}
Here, $\Xi_j=\frac{1}{3}i \mu m\omega_j^2$ $(j=1,2,3)$, $\alpha_j=\left\langle a_j \right\rangle$ $(j=1,2)$, and $\beta_j=\left\langle b_j \right\rangle$ $(j=1,2,3)$. For the convenience of discussion, we assume $\omega_j=\omega_m$ $(j=1,2,3)$.
%figure 6
\begin{figure}[h]
\centering
\includegraphics[width=0.50\textwidth]{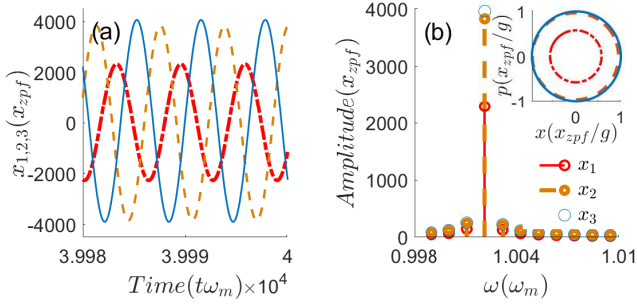}
\caption {(a) The time evolution of the three resonators with $\alpha^{in}=160\sqrt{\omega_m}$. Red dash-dotted, orange dashed, and blue solid lines for different resonators. (b) Fourier spectra and its corresponding phase space trajectories. The coupling has been fixed at $J = 2.2\times10^{-3}\omega_m$, and the other parameters are the same as in Fig. \ref{twostate_benzhengzhi}.}
	\label{threestate_wentai_FFT}
\end{figure}

In Fig. \ref{threestate_wentai_FFT}, we show the overall properties of the steady state solutions of the mechanical resonators. It is easy to find that the amplitudes of the the mechanical resonators change  very slowly over time [see Fig. \ref{threestate_wentai_FFT} (a)]. Fig. \ref{threestate_wentai_FFT} (b) shows the corresponding Fourier spectra. It is easy to see that all three mechanical resonators start oscillating with a same frequency, i.e., $\omega_l=\frac{\omega_{eff}^1+\omega_2+\omega_{eff}^3}{3}$. The inset of Fig. \ref{threestate_wentai_FFT} (b) shows limit cycle oscillations at $\alpha^{in}=160\sqrt{\omega_m}$ and $J=2.2\times10^{-3}\omega_m$. So in this case, the formal solution for $\beta_j(t)$ is still applicable. By the use of  this formal solution, Eq. (\ref{QLEs-EP3}) can be further reduced to
\begin{small}
\begin{eqnarray}
\begin{aligned}
&\frac{d\beta_1}{dt}=-(i\omega_{eff}^1+\frac{\gamma_{eff}^1}{2}+i \Theta_1)\beta_1+iJ\beta_{2}+i \Theta_1\beta_1^*,
\\&\frac{d\beta_2}{dt}=-(i\omega_2+\frac{\gamma_m}{2}+i \Theta_2)\beta_2+iJ\beta_{1}+iJ\beta_{3}+i \Theta_2\beta_2^*,
\\&\frac{d\beta_3}{dt}=-(i\omega_{eff}^3+\frac{\gamma_{eff}^3}{2}+i \Theta_3)\beta_3+iJ\beta_{2}+i \Theta_3\beta_3^*.
\label{QLEs-EP3_reduce}
\end{aligned}
\end{eqnarray}
\end{small}
Here $\Theta_j= \mu m \omega_j^2 B_j^2$ $(j=1,2,3)$. The modal field evolution in this configuration obeys $i\frac{d\psi}{dt} =  H_{eff} \psi$, where $\psi = (\beta_1, \beta_2,\beta_3,\beta_1^*,\beta_2^*,\beta_3^*)^T$ represents the modal state vector and $t$ represents time. $H_{eff}$ is the associated $6\times6$ non-Hermitian Hamiltonian,

\begin{small}
\begin{widetext}
\begin{equation}
\left(
 \begin{array}{cccccc}
\omega_{eff}^1-i\frac{\gamma_{eff}^1}{2}+\Theta_1&-J&0&-\Theta_1&0&0\\
 -J&\omega_2-i\frac{\gamma_m}{2}+\Theta_2&-J&0&-\Theta_2&0\\
 0&-J&\omega_{eff}^3-i\frac{\gamma_{eff}^3}{2}+\Theta_3&0&0&-\Theta_3\\
 \Theta_1&0&0&-\omega_{eff}^1-i\frac{\gamma_{eff}^1}{2}-\Theta_1&J&0\\
 0&\Theta_2&0&J&-\omega_2-i\frac{\gamma_m}{2}-\Theta_2&J\\
 0&0&\Theta_3&0&J&-\omega_{eff}^3-i\frac{\gamma_{eff}^3}{2}-\Theta_3
 \end{array}
\right).
\end{equation}
\end{widetext}
\end{small}

It is easy to find that the effective Hamiltonian has 6 eigenvalues forming two pairs, one pair is due to the apperarnace of $\beta_j^*$ in the dynamics.
    %figure 7
\begin{figure}[h]
	\centering
	\includegraphics[width=0.50\textwidth]{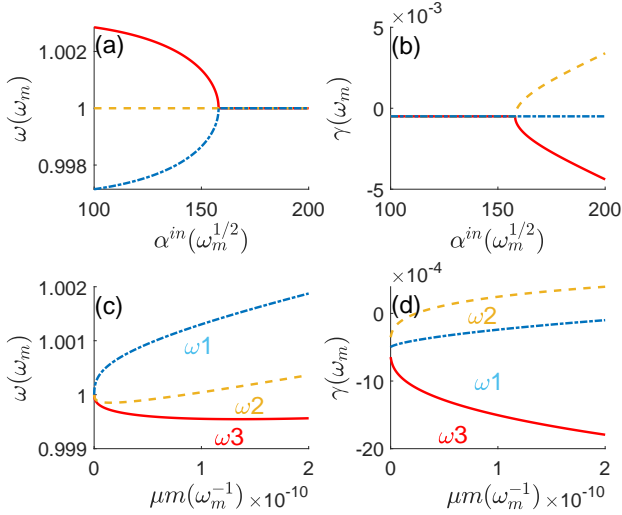}
	\caption {(a) The real and (b) the imaginary parts of the eigenvalues as a function of the driving strength $\alpha^{in}$ before the perturbation introduced by gravitational effects, where the corresponding mechanical coupling strength $J = 2.2\times10^{-3}\omega_m$. (c)  The real and (d) the imaginary parts of the eigenvalues of the ternary mechanical system as a function of $\mu m$ around a third-order exceptional point for a fixed $J = 2.2\times10^{-3}\omega_m$. The three eigenvalues of the effective Hamiltonian are marked with red solid, orange dashed, and blue dash-dotted lines. The other system parameters are the same as in Fig. \ref{twostate_benzhengzhi}.}
	\label{threestate_benzhengzhi}
\end{figure}
This characteristic feature of the EP3 has been demonstrated in Fig. \ref{threestate_benzhengzhi} (a) and (b), where we show the dependence of the eigenvalues on driving strength $\alpha^{in}$.

Now to take this discussion further to show  how the system reacts around the EP3. The real [Fig. \ref{threestate_benzhengzhi} (c)] and imaginary parts [Fig. \ref{threestate_benzhengzhi} (d)] of the eigenvalues are  plotted as a function of $\mu m$. Moreover, it is easy to see that the power $(p_{in})$ required to reach the third-order exceptional point is lower than that required by EP2.
    %figure 8
\begin{figure}[h]
	\centering
	\includegraphics[width=0.5\textwidth]{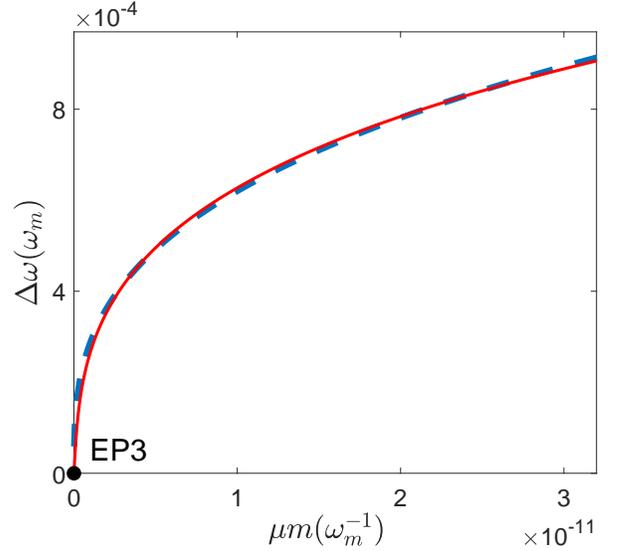}
	\caption {Frequency splitting (red solid line) in the mechanical supermodes $vs$ $\mu m$ near EP3 for $J = 2.2\times10^{-3}\omega_m$. The blue dashed line denotes the fitted curve with $\varsigma= 2.874\omega_m^{4/3}$. The other parameters are the same as in Fig. \ref{twostate_benzhengzhi}.}
	\label{lingmindu_EP3}
\end{figure}

The difference between two effective frequencies (in this case, $\omega_2$ and $\omega_3$) is also plotted (Fig. \ref{lingmindu_EP3}) as a function of $\mu m$. The
frequency splitting caused by $\mu m$ can be fitted using
\begin{eqnarray}
\begin{aligned}
\Delta\omega\approx\varsigma (\mu m)^{1/3}.
\label{fit_EP3}
\end{aligned}
\end{eqnarray}
Here $\varsigma$ is the fitting coefficient. The blue dashed line represents the fitting results  according to Eq. (\ref{fit_EP3}) with $\varsigma= 2.874\omega_m^{4/3}$, which is consistent with the mechanical frequency splitting in our system,  confirming thus that the mechanical frequency splitting in response to $\mu m$ obeys the cube root behavior. This indicates that it is feasible to further enhance the sensitivity by means of third-order exceptional point (EP3).

\section{Experimental feasibility and ultimate limits  of the  sensing scheme}\label{sec4}

\begin{figure}[h]
	\centering
	\includegraphics[width=0.44\textwidth]{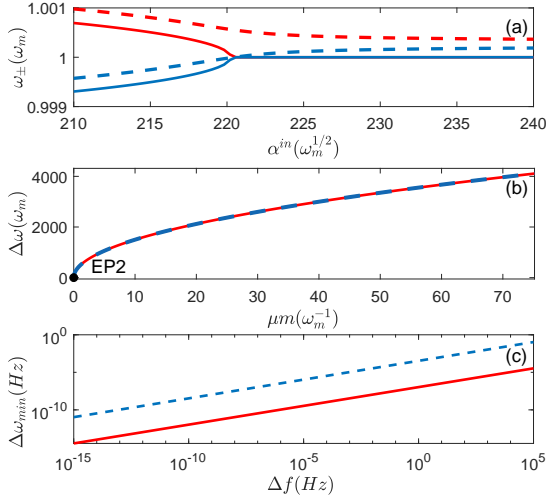}
	\caption {(a) The real parts of the eigenvalues $vs$ the driving strength $\alpha^{in}$ before perturbation (solid lines) and after perturbation (dashed lines) by gravitational effects with $\mu m=(0.02\times10^{-19})\omega_m^{-1}$. The two eigenvalues of our model are marked with red and blue lines. (b) Frequency splitting (red solid line) in the mechanical supermodes $vs$ $\mu m$ near EP2. The blue dashed line denotes the fitted line, where the fitting coefficient is $\xi=474.1 \omega_m^{3/2}$. (c) Limit of the sensitivity of the system to gravitational effects imposed by thermomechanical noise as a function of the bandwidth with $Q=10^{12}$ (red solid line) and $Q=10^5$ (blue dashed line) in the condition of $T=300K$. The experimentally realistic parameters are chosen as $\omega_m/2 \pi = 6$ GHz, $\gamma_m/2 \pi=3$ MHz, $g/2 \pi=0.8$ MHz, $\kappa/2 \pi=2$ GHz, and $J/2 \pi=10$ MHz \cite{Fang2016}.}
	\label{Experimental feasibility}
\end{figure}

There are many types of optomechanical systems. For concreteness, we choose one of them, where the mechanical degree of freedom is a dielectric membrane placed inside a Fabry-Perot cavity \cite{Thompson2008}. Here we use two coupled Si beams, which possess the mass of $m = 5.3\times10^{-3}$ ng and thickness $t = 80$ nm \cite{Ekincia2004}. Here we take the EP2-based sensor as an example, as shown in Fig. \ref{Experimental feasibility} (a) and (b). In general, various basic physical noise processes will limit the sensitivity of the  sensing scheme. For the nanomechanical resonators, the main noise source is the thermomechanical noise \cite{Ekincia2004}. In order to obtain this basic limits imposed upon measurements by thermomechanical fluctuations, we need to consider the minimum detectable frequency shift $(\delta\omega)$ that can be resolved in a practical noisy system. An estimate for $\delta\omega$ can be obtained by \cite{Ekincia2004}
\begin{eqnarray}
\begin{aligned}
\delta\omega\approx\big(\frac{K_B T}{E_c} \frac{\omega_n \Delta f}{Q}\big)^{1/2}.
\label{quantum-noise-limited sensitivity}
\end{aligned}
\end{eqnarray}
Here $Q$ is the mechanical quality factor, $K_B$ is the Boltzmann constant, $T$ is the effective temperature of the mechanical resonator, and $E_c=m \omega_m^2 \left\langle x_c^2\right\rangle$, which describes the maximum drive energy. $\left\langle x_c\right\rangle$ can be approximated as \cite{Ekincia2004}
\begin{eqnarray}
\begin{aligned}
\left\langle x_c\right\rangle \approx 0.53 t.
\label{x_c}
\end{aligned}
\end{eqnarray}
In order to obtain the ultimate sensitivity limits of the system to the effect of gravity, we assume that the frequency splitting caused by gravitational effects is exactly equal to the minimum measurable frequency shift $(\delta\omega)$ determined by the thermomechanical fluctuations, i.e., $\Delta\omega_{min} = \delta \omega$. We plot $\Delta\omega_{min}$ as a function of the bandwidth for thermomechanical fluctuations in Fig. \ref{Experimental feasibility} (c). The result shows that  small bandwidth $\Delta f$ and  high quality factor $Q$ of the mechanical resonator are essential for the superresolution. Assuming that $\Delta f=10^{-10}$ Hz, we can obtain the quantum-noise-limited sensitivity of the system to gravitational effects with Eq.(\ref{quantum-noise-limited sensitivity}), $\Delta\omega_{min}\sim10^{-12}$ Hz for $Q=10^{12}$.
    %figure 9

  \section{Conclusion}\label{sec5}
In conclusion, we  have presented a scheme for  sensing the effect of quantum  gravity. Starting with a system consisting  of two coupled resonators with driving and dissipation,  we show that the system eigenenergy is sensitive to the effect of  quantum gravity  when the system is  in an second-order exceptional point. The response of the binary mechanical system to the gravity exhibits square root behaviour, and the sensitivity of the system at EPs increases significantly with the decrease of the perturbation. Moreover, we found that small mass of the mechanical resonator benefits  the sensitivity of the system to deposition mass. In order to further enhance the sensitivity of the system to   the effect of  gravity,  we extend the sensing scheme  to a third-order exceptional point by taking a more complicated ternary mechanical system into account. The response of the ternary mechanical systems to perturbation exhibits cube root behaviour. The quantum-noise-limited sensitivity of the system to  gravitational effects due to thermomechanical noise is also discussed. It is  worthwhile to note that our scheme could, in principle, be extended to various photonic and phononic systems with optomechanically induced gain and loss. These findings may pave the ways for utilizing EPs as a novel tool to probe effect of quantum  gravity.
  \section{acknowledgments}\label{sec6}
This work is supported by National Natural Science Foundation of China (NSFC) under Grants No. $11775048$ and No. $11947405$.

\begin{appendix}
\section{the derivation of mechanical effective Hamiltonian}\label{appendix}
Based on this  formal solution: $\beta_j(t)=\bar{\beta}_j +B_je^{-i\theta} e^{-i\omega_{l} t}$ $(\bar{\beta}_j\ll B_j)$ \cite{Marquardt2006,Rodrigues2010}, Eq. (\ref{QLEs1}) can be further simplified as
\begin{small}
\begin{eqnarray}
\begin{aligned}
&\frac{d\beta_1}{dt}=-(i\omega_1+\frac{\gamma_m}{2})\beta_1+iJ\beta_2+ig\alpha_1^* \alpha_1+i \Theta_1(-\beta_1+\beta_1^*),
\\&\frac{d\beta_2}{dt}=-(i\omega_2+\frac{\gamma_m}{2})\beta_2+iJ\beta_1+ig\alpha_2^* \alpha_2+i \Theta_2(-\beta_2+\beta_2^*),
%\\&\frac{d\beta_1^*}{dt}=-(-i\omega_1+\frac{\gamma_m}{2})\beta_1^*-iJ\beta_2^*-ig\alpha_1^* \alpha_1-i\Theta_1(-\beta_1^*+\beta_1),
%\\&\frac{d\beta_2^*}{dt}=-(-i\omega_2+\frac{\gamma_m}{2})\beta_2^*-iJ\beta_1^*-ig\alpha_2^* \alpha_2-i\Theta_2(-\beta_2^*+\beta_2),
\label{QLEsa1}
\end{aligned}
\end{eqnarray}
\end{small}
where, $\Theta_j= \mu m \omega_j^2 B_j^2 (j=1,2)$. We substitute this formal solution into the equation for $\alpha_j$, one then obtain the dynamics of the cavity field in the form,
\begin{eqnarray}
\begin{aligned}
\alpha_j(t)=exp(-i\varphi_j(t))\sum_{n}A_n^j exp(in\omega_{l}t),
\label{the dynamics of the cavity field}
\end{aligned}
\end{eqnarray}
with
\begin{eqnarray}
\begin{aligned}
A_n^j=\sqrt{\kappa}\alpha^{in}\frac{J_n(-\epsilon_j)}{K_n^j},
\label{amplitude}
\end{aligned}
\end{eqnarray}
where $\epsilon_j=2g B_j/\omega_{l}$ is normalized amplitude, $\Delta_j^{'}=\Delta_j+2gRe(\bar{\beta}_j)$, $K_n^j=i(n\omega_{l}-\Delta_j^{'})+\frac{\kappa}{2}$, the global phase is $\varphi(t)=-\epsilon_j sin(\omega_{l} t-\theta)$ and $J_n$ is the Bessel function of the first kind.

As we pay our attention to the limit-cycle states of the mechanical resonators, we removed all terms in mechanical dynamics except for the constant one and the term oscillating at $\omega_l$. We substitute Eq. (\ref{the dynamics of the cavity field}) into Eq. (\ref{QLEsa1}) which leads to the following equations of motion for the oscillating part of $\beta_j$ $(\bar{\beta}_j\ll B_j)$,
\begin{small}
\begin{eqnarray}
\begin{aligned}
&\frac{d\beta_1}{dt}=-(i\omega_{eff}^1+\frac{\gamma_{eff}^1}{2}+i \Theta_1)\beta_1+iJ\beta_2+i \Theta_1 \beta_1^*,
\\&\frac{d\beta_2}{dt}=-(i\omega_{eff}^2+\frac{\gamma_{eff}^2}{2}+i \Theta_2)\beta_2+iJ\beta_1+i \Theta_2 \beta_2^*,
%\\&\frac{d\beta_1^*}{dt}=-(-i\omega_{eff}^1+\frac{\gamma_{eff}^1}{2})\beta_1^*-iJ\beta_2^*-i\Theta_1(-\beta_1^*+\beta_1),
%\\&\frac{d\beta_2^*}{dt}=-(-i\omega_{eff}^2+\frac{\gamma_{eff}^2}{2})\beta_2^*-iJ\beta_1^*-i\Theta_2(-\beta_2^*+\beta_2),
\label{QLEsa4}
\end{aligned}
\end{eqnarray}
\end{small}
where, $\omega_{eff}^j=\omega_j+\Omega_j$ and $\gamma_{eff}^j=\gamma_m+\Gamma_j$ $(j=1,2)$ represent the effective frequency and the effective damping of the $j$th mechanical oscillator $(j = 1, 2)$, respectively. The optical spring effect $(\Omega_j)$ and optomechanical damping rate $(\Gamma_j)$ of the mechanical resonator due to the cavity are given by \cite{Rodrigues2010}
\begin{eqnarray}
\begin{aligned}
\Omega_j&=-\frac{2\kappa(g\alpha^{in})^2}{\omega_{l} \epsilon_j} Re\left(\sum_{n}\frac{J_{n+1}(-\epsilon_j)J_{n}(-\epsilon_j)}{K_{n+1}^{j*}K_n^j}\right),
\label{optical spring effect_a5}
\end{aligned}
\end{eqnarray}
and
\begin{eqnarray}
\begin{aligned}
\Gamma_j&=\frac{2(g\kappa\alpha^{in})^2}{\epsilon_j} \sum_{n}\frac{J_{n+1}(-\epsilon_j)J_{n}(-\epsilon_j)}{\left|K_{n+1}^{j*}K_n^j\right|^2}.
\label{optomechanical damping rate_a6}
\end{aligned}
\end{eqnarray}
Here, $\Omega_j$ $(\Gamma_j)$ can be controlled by the external drive signal. If the optomechanical system satisfies the resolved-sideband condition, $\gamma_m,g\ll\kappa\ll\omega_m$ \cite{Cohen2015,Hong2017}, the optical spring effect can be ignored. Further, the effective Hamiltonian of the mechanical modes can be derived as
\begin{widetext}
\begin{equation}
H_{eff}= \left(
 \begin{array}{cccc}
\omega_{eff}^1-i\frac{\gamma_{eff}^1}{2}+\Theta_1&-J&-\Theta_1&0\\
 -J&\omega_{eff}^2-i\frac{\gamma_{eff}^2}{2}+\Theta_2&0&-\Theta_2\\
 \Theta_1&0&-\omega_{eff}^1-i\frac{\gamma_{eff}^1}{2}-\Theta_1&J\\
 0&\Theta_2&J&-\omega_{eff}^2-i\frac{\gamma_{eff}^2}{2}-\Theta_2
 \end{array}
\right).
\end{equation}
\end{widetext}
\end{appendix}

\end{document}